\documentclass{emulateapj}
\usepackage{apjfonts}
\lefthead{HAN} 
\righthead{MICROLENSING FOLLOW-UP CRITERIA}

\begin{document}
\title{Criteria in the Selection of Target Events for Planetary Microlensing 
Follow-Up Observations}

\author{Cheongho Han}
\affil{
Program of Brain Korea 21,
Institute for Basic Science Research, 
Department of Physics,\\
Chungbuk National University, Chongju 361-763, Korea;
cheongho@astroph.chungbuk.ac.kr}



\begin{abstract}

To provide criteria in the selection of target events preferable for 
planetary lensing follow-up observations, we investigate the variation 
of the probability of detecting planetary signals depending on the 
observables of the lensing magnification and source brightness.   
In estimating the probability, we consider variation of the photometric 
precision by using a quantity defined as the ratio of the fractional 
deviation of the planetary perturbation to the photometric precision.  
From this investigation, we find consistent result from previous studies 
that the probability increases with the increase of the magnification.  
The increase rate  is boosted at a certain magnification at which 
perturbations caused by central caustic begin to occur.  We find this 
boost occurs at moderate magnifications of $A\lesssim 20$, implying that 
probability can be high even for events with moderate magnifications.  
The probability increases as the source brightness increases.  We find 
that the probability of events associated with stars brighter than clump 
giants is not negligible even at magnifications as low as $A\sim 5$.  
In the absence of rare the prime target of very high-magnification events, 
we, therefore, recommend to observe events with brightest source stars 
and highest magnifications among the alerted events.  Due to the increase 
of the source size with the increase of the brightness, however, the 
probability rapidly drops off beyond a certain magnification, causing 
detections of low mass ratio planets ($q\lesssim 10^{-4}$) difficult from 
the observations of events involved with giant stars with magnifications 
$A\gtrsim 70$.
\end{abstract}


\section{Introduction}

With the advantages of being able to detect very low-mass planets and 
those with separations from host stars that cannot be covered by other 
methods, microlensing is one of the most important methods that can 
detect and characterize extrasolar planets \citep{mao91, gould92}.  The 
microlensing planetary signal is a short duration perturbation to the 
standard lensing light curve produced by the primary star.  To achieve 
high monitoring frequency required for the detection of the short-lived 
planetary signal, current lensing experiments are employing early-warning 
system to issue alerts of ongoing events in the early stage of lensing 
magnification \citep{udalski94, bond02} and follow-up observations to 
intensively monitor the alerted events \citep{dominik02, yoo04}.  Under 
current surveys, there exist in average $\gtrsim 50$ alerted events at 
a certain time \citep{dominik02}.  Then, an important issue related to 
the follow-up observation is which event should be monitored for better 
chance of planet detections.

There have been several estimates of microlensing planet detection 
efficiencies \citep{bolatto94, bennett96, gaudi00, peale01}.   Most of 
these works estimated the efficiency as a function of the instantaneous 
angular star-planet separation normalized by the angular Einstein radius, 
$s$, and planet/star mass ratio, $q$.  However, the efficiency determined 
in this way is of little use in the point of view of observers who are 
actually carrying out follow-up observations of lensing events.  This is 
because the planet parameters $s$ and $q$ are not known in the middle of
lensing magnification and thus they cannot be used as criteria in the 
selection of target events for follow-up observations.  Related to the 
target selection, \citet{griest98} proposed a useful criterion to observers.  
They pointed out that by focusing on very high-magnification ($A\gtrsim 100$) 
events, the probability of detecting planets in the lensing zone could be 
very high.  However, these events are rare and thus they cannot be usually 
found in the list of alerted events.  Therefore, it is necessary to have 
criteria applicable to general lensing events in the absence of very 
high-magnification events.  To provide such criteria, we investigate 
the dependency of the probability of detecting planetary signals on the 
observables such as the lensing magnification and source type.

The paper is organized as follows.  In \S\ 2, we briefly describe the 
basics of planetary microlensing.  In \S\ 3, we investigate the variation 
of the probability of detecting planetary signals depending on the lensing 
magnification and source type for events caused by planetary systems with 
different masses and separations.  We analyze the result and qualitatively 
explain the tendencies found from the investigation.  Based on the result 
of the investigation, we then present criteria for the selection of target 
events preferable for follow-up observations.  In \S\ 4, we summarize the 
results and conclude.

\section{Basics of Planetary Lensing}

The lensing behavior of a planetary lens system is described by the 
formalism of a binary lens with a very low-mass companion.  Because of 
the very small mass ratio, planetary a lensing light curve is well 
described by that of a single lens of the primary star for most of the 
event duration.  However, a short-duration perturbation can occur when 
the source star passes the region around the caustics, that are the set 
of source positions at which the magnification of a point source becomes 
infinite.  The caustics of binary lensing form a single or multiple sets 
of closed curves where each of which is composed of concave curves (fold 
caustics) that meet at points (cusps).

For a planetary case, there exist two sets of disconnected caustics: 
`central' and `planetary' caustics.  The single central caustic is located 
close to the host star.  It has a wedge shape with four cusps and its size 
(width along the star-planet axis) is related to the planet parameters by 
\citep{chung05}
\begin{equation}
\Delta\xi_{\rm cc} \propto {q\over (s-1/s)^2}.
\label{eq1}
\end{equation}
For a given mass ratio, a pair of central caustics with separations $s$ 
and $s^{-1}$ are identical to the first order of approximation 
\citep{dominik99, griest98,an05}.  The planetary caustic is located away 
from the host star.  The center of the planetary caustic is located on 
the star-planet axis and the position vector to the center of the planetary 
caustic measured from the primary lens position is related to the lens-source 
separation vector, ${\bf s}$, by 
\begin{equation}
{\bf r}_{\rm pc}={\bf s}\left(1-{1 \over s^2}\right).
\label{eq2}
\end{equation}
Then, the planetary caustic is located on the planet side, i.e.\ ${\rm sign} 
({\bf r}_{\rm pc})= {\rm sign}({\bf s})$, when $s>1$, and on the opposite 
side, i.e.\ ${\rm sign} ({\bf r}_{\rm pc})=-{\rm sign}({\bf s})$, when 
$s<1$.  When $s>1$, there exists a single planetary caustic and it has a 
diamond shape with four cusps.  When $s<1$, there are two caustics and 
each caustic has a triangular shape with three cusps.  The size of the 
planetary caustic is related to the planet parameters by
\begin{equation}
\Delta\xi_{\rm pc} \propto 
\cases{
q^{1/2}/(s\sqrt{s^2-1})                                & for $s > 1$,\cr
q^{1/2}(\kappa_0-1/\kappa_0+\kappa_0/s^2)\cos\theta_0  & for $s < 1$,\cr
}
\label{eq3}
\end{equation}
where $\kappa (\theta) = \left\{[\cos 2\theta\pm (s^4-\sin^2 2\theta)^{1/2}]
/ (s^2-1/s^2) \right\}^{1/2}$, $\theta_0 = [\pi \pm \sin^{-1}(3^{1/2}s^2/2)]
/2$, and $\kappa_0=\kappa(\theta_0)$ \citep{han06}.  The planetary caustic 
is always bigger than the central caustic and the size ratio between the 
two types of caustics, $\Delta\xi_{\rm cc}/\Delta \xi_{\rm pc}$, becomes 
smaller as the mass of the planet becomes smaller and the planet is located 
further away from the Einstein ring.  The planetary caustic is located 
within the Einstein ring of the primary when the planet is located in 
the range of separation from the star of $0.6\lesssim s\lesssim 1.6$.  
The size of the caustic, which is directly proportional to the planet 
detection efficiency, is maximized when the planet is located in this 
range, and thus this range is called as the `lensing zone'.  As the position 
of the planet approaches to the Einstein ring radius, $s\rightarrow 1$, 
the location of the planetary caustic approaches the position of the 
central caustic. Then, the two types of caustic eventually merge together, 
forming a single large one.

\section{Variation of Detectability}

\subsection{Quantification of Detectability}

The quantity that has been often used in the previous estimation of the 
planet detection probability is the `fractional deviation' of the planetary 
lensing light curve from that of the single lensing event of the primary, 
i.e.,
\begin{equation}
\epsilon = {A-A_0 \over A_0}.
\label{eq4}
\end{equation}
With this quantity, however, one cannot consider the variation of the 
photometric precision depending on the lensing magnification.  In addition, 
it is difficult to consider the variation of the detectability depending 
on the source type.

To consider the effect of source star brightness and its lensing-induced 
variation on the planet detection probability, we carry out our analysis 
based on a new quantity defined as the ratio of the fractional deviation, 
$\epsilon$, to the photometric precision, $\sigma_\nu$, i.e, 
\begin{equation}
{\cal D}={\left\vert \epsilon\right\vert \over \sigma_\nu};\qquad
\sigma_\nu = { ( AF_{\nu,{\rm S}}+F_{\nu,{\rm B}})^{1/2} 
\over (A-1)F_{\nu,{\rm S}}},
\label{eq5}
\end{equation}
where $F_{\nu,{\rm S}}$ 
and $F_{\nu,{\rm B}}$ represent the fluxes from the source star and 
blended background stars, respectively.  Here we assume that photometry 
is carried out by using the difference imaging method \citep{tomaney96, 
alard99}.  In this technique, photometry of the lensed source star is 
conducted on the subtracted image obtained by convolving two images taken 
at different times after geometrically and photometrically aligning them.  
Then the signal from the lensed star measured on the subtracted image is 
the flux variation of the lensed source star, $(A-1) F_{\nu,{\rm S}}$, 
while the noise originates from both the source and background blended 
stars, $ AF_{\nu, {\rm S}}+ F_{\nu,{\rm B}}$.  Under this definition of 
the planetary signal detectability, ${\cal D}=1$ implies that the planetary 
signal is equivalent to the photometric precision.  Hereafter we refer the 
quantity ${\cal D}$ as the `detectability'.

\begin{figure*}[t]
\epsscale{1.0}
\caption{\label{fig:one}
Contour maps of the detectability of the planetary signal, ${\cal D}$, 
as a function of the position in the source plane for events caused by 
planetary systems with various lens-source separations and mass ratios.  
The detectability represents the ratio of the fractional deviation of 
the planetary lensing light curve from the single lensing light curve of 
the primary to the photometric precision.  All lengths are normalized by 
the angular Einstein radius and $\xi$ and $\eta$ represent the coordinates 
parallel with and normal to the star-planet axis, respectively.  The 
individual sets of panels show the maps for events associated with different 
types of source stars.  Contours (yellow curve) are drawn at the level 
of ${\cal D}=3.0$.  The maps are centered at the position of the primary 
lens star and the planet is located on the left.  The dotted arc in each 
panel represents the Einstein ring of the primary star.  The closed figures 
drawn by red curves represent the caustics.  For the details about the 
assumed lens parameters and observational conditions, see \S\ 3.2.
}\end{figure*}

\subsection{Contour Maps of Detectability}

To see the variation of the detectability depending on the separation 
parameter $s$, mass ratio $q$, and the types of involved source star, we 
construct maps of detectability as a function of the position in the 
source plane.  Figure~\ref{fig:one} shows example maps.  The individual 
sets of panels show the maps for events associated with different types 
of source stars.  All lengths are normalized by the angular Einstein radius 
and $\xi$ and $\eta$ represent the coordinates parallel with and normal 
to the star-planet axis, respectively.  A contours (yellow curve) is drawn 
at the level of ${\cal D}=3.0$.  The maps are centered at the position of 
the primary lens star and the planet is located on the left.  The dotted 
arc in each panel represents the Einstein ring of the primary star.  The 
closed figures drawn by red curves represent the caustics.

For the construction of the maps, we assume a mass of the primary lens 
star of $m=0.3\ M_\odot$ and distances to the lens and source of $D_{\rm L}
=6$ kpc and $D_{\rm S}=8$ kpc, respectively.  Then, the corresponding 
Einstein radius is $r_{\rm E}=\left\{ (4Gm/c^2) [(D_{\rm L}(D_{\rm S}-
D_{\rm L})/D_{\rm S}] \right\}^{1/2}=1.9$ AU.  For the source stars, we 
test three different types of giant, clump giant, and main-sequence stars.  
The assumed $I$-band absolute magnitudes of the individual types of stars 
are $M_I=0.0$, 1.5, and 3.6, respectively.  With the assumed amount of 
extinction toward the Galactic bulge field of $A_I=1.0$, these correspond 
to the apparent magnitudes of $I=15.5$, 17, and 19.1, respectively.  As 
the source type changes, not only the brightness but also the size of the 
star changes.  Source size affects the planetary signal in lensing light 
curves \citep{bennett96} and thus we take account the finite source effect 
into consideration.  The assumed source radii of the individual types of 
source stars are 10.0 $R_\odot$, 3.0 $R_\odot$, and 1.1 $R_\odot$, 
respectively.  We assume that events are affected by blended flux equivalent 
to that of a star with $I=20$.  We note that the adopted lens and source 
parameters are the typical values of Galactic bulge events that are being 
detected by the current lensing surveys \citep{han03}.

For the observational condition, we assume that images are obtained by 
using 1 m telescopes, which are typical ones being used in current 
follow-up observations.  We also assume that the photon acquisition 
rate of each telescope is 10 photons per second for an $I=20$ star and a 
combined image with a total exposure time of 5 minutes is obtained from 
each set of observations.

\begin{figure}[t]
\epsscale{1.25}
\plotone{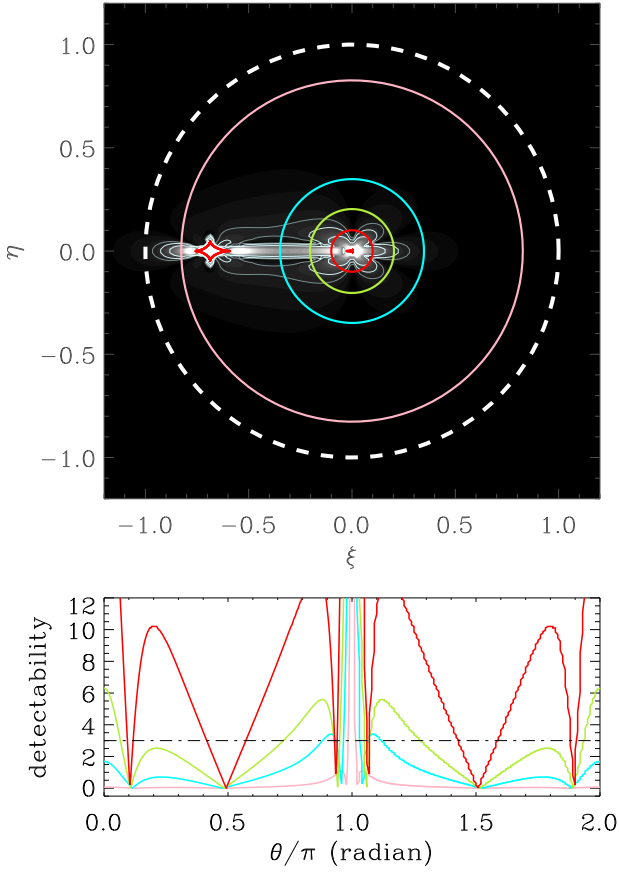}
\caption{\label{fig:two}
Geometric representation of the probability of detecting planetary signals, 
$P$.  Under the definition of $P$ as {\it the average probability of 
detecting planetary signals with a detectability greater than a threshold 
value $D_{\rm th}$ at the time of observation with a magnification $A$}, 
the probability corresponds to the portion of the arclet(s) where the 
detectability is greater than a threshold value out of a circle around 
the primary with a radius equal to the lens-source separation corresponding 
to the magnification at the time of observation.  The individual circles 
in the upper panel correspond to the source positions at which the lensing 
magnifications are $A=1.5$ (pink), 3.0 (cyan), 5.0 (green), and 10.0 (red), 
respectively.  The curves in the bottom panels show the variation of the 
detectability as a function of the position angle ($\theta$) of points on 
the circles with corresponding colors in the upper panel.  We set the 
threshold detectability as ${\cal D}_{\rm th}=3.0$, i.e.\ $3\sigma$ 
detection of the planetary signal.  The dashed circle represents the 
Einstein ring.
}\end{figure}

\begin{figure}[t]
\epsscale{1.2}
\plotone{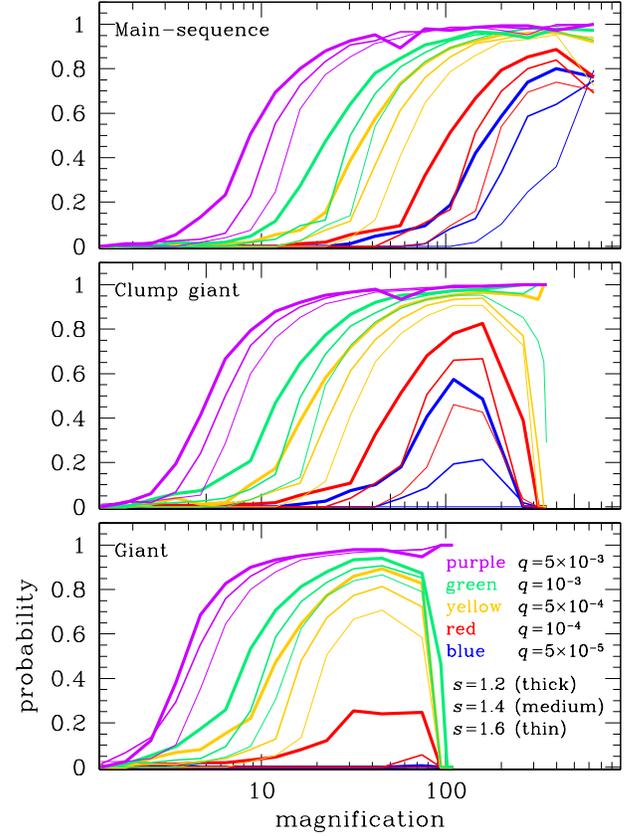}
\caption{\label{fig:three}
Probability of detecting planetary signals as a function of lensing 
magnification.  The individual panels show the probabilities for events 
involved with different types of source stars.  The curves in each panel 
show the variation of the probability for planets with different mass 
ratios and separations.  We note that although not presented, the 
probabilities for planets with separations $s<1$ are similar to those of 
the corresponding planets with $s^{-1}$.  The probability is defined the 
average probability of detecting planetary signals with a detectability 
greater than a threshold value $D_{\rm th}$ at the time of observation 
with a magnification $A$.  We set the threshold detectability as 
${\cal D}_{\rm th}=3.0$, i.e.\ $3\sigma$ detection of the planetary signal.
We note that there is a maximum magnification specific to the angular size 
of the source star and thus the curves stop at certain magnifications.
}\end{figure}

\subsection{Probability of Detecting Planetary Signals}

Based on the maps of detectability, we then investigate the probability
of detecting planetary signals as a function of the lensing magnification.  
We define the probability $P$ as {\it the average probability of detecting 
planetary signals with a detectability greater than a threshold value 
$D_{\rm th}$ at the time of observation with a magnification $A$}.  
Geometrically, this probability corresponds to the portion of the arclet(s) 
where the detectability is greater than a threshold value out of a circle 
around the primary with a radius equal to the lens-source separation 
corresponding to the magnification at the time of observation.  This is 
illustrated in Figure~\ref{fig:two}.  We note that the magnification is 
a unique function of the absolute value of the lens-source separation 
$u$\footnote{Strictly speaking, the magnification depends additionally 
on the size of the source star.}, and thus $A={\rm const}$ corresponds 
to a circle around the lens.  The lens-source separation is related to 
the magnification by 
\begin{equation}
u(A) = 
\left[ {2\over (1-A^{-2})^{1/2}}-2 \right]^{1/2}.  
\label{eq6}
\end{equation}
We set the threshold detectability as $D_{\rm th}=3.0$, i.e.\ $3\sigma$ 
detection of the planetary signal.

In Figure~\ref{fig:three}, we present the resulting probability as a 
function of magnification.  The individual panels show the probabilities 
for events involved with different types of source stars.  In each panel, 
we present the variations of the probability for planets with different 
mass ratios and separations.  We test six different planetary separations 
of $s=1/1.6$, 1/1.4, 1/1.2, 1.2, 1.4, and 1.6 as representative values 
for planets in the lensing zone.  For the mass ratio, we test five values 
of $q=5\times 10^{-3}$, $10^{-3}$, $5\times 10^{-4}$, $10^{-4}$, 
and $5\times 10^{-5}$.

\begin{deluxetable}{ll}
\tablecaption{Limitation by Finite-Source Effect\label{table:one}}
\tablewidth{0pt}
\tablehead{
\colhead{source type} &
\colhead{event type} 
}
\startdata
giant         & $A\gtrsim 70$ for planets with $q\lesssim 10^{-3}$ \\
clump giant   & $A\gtrsim 200$ for planets with $q\lesssim 5\times 10^{-4}$ \\
main-sequence & $A\gtrsim 500$ for planets with $q\lesssim 10^{-4}$ \\
\enddata 
\tablecomments{ 
Cases of planetary microlensing events where detection of planetary 
signal is limited by finite source effect.  We note that ``-'' means 
the respective configuration cannot be realized. 
}
\end{deluxetable}

\begin{deluxetable*}{llccccc}
\tablecaption{Critical Magnifications of Central Perturbation\label{table:two}}
\tablewidth{0pt}
\tablehead{
\multicolumn{1}{l}{source} &
\multicolumn{1}{c}{planetary} &
\multicolumn{5}{c}{mass ratio} \\
\multicolumn{1}{l}{type} &
\multicolumn{1}{c}{separation} &
\multicolumn{1}{c}{$q=5\times 10^{-3}$} &
\multicolumn{1}{c}{$q=10^{-3}$} &
\multicolumn{1}{c}{$q=5\times 10^{-4}$} &
\multicolumn{1}{c}{$q=10^{-4}$} &
\multicolumn{1}{c}{$q=5\times 10^{-5}$} 
}
\startdata
              & $s=1.2$, 1/1.2  & $A\sim 2.2$  & $A\sim 7$  & $A\sim 8$   & $A\sim 22$     & $A\sim 22$  \\
giant         & $s=1.4$, 1/1.4  & $A\sim 2.5$  & $A\sim 8$  & $A\sim 12$  & --             & --          \\
\smallskip
              & $s=1.6$, 1/1.6  & $A\sim 3.5$  & $A\sim 9$  & $A\sim 18$  & --             & --          \\

clump         & $s=1.2$, 1/1.2  & $A\sim 7$    & $A\sim 8$    & $A\sim 11$  & $A\sim 30$   & $A\sim 60$  \\
giant         & $s=1.4$, 1/1.4  & $A\sim 8$    & $A\sim 12$   & $A\sim 17$  & $A\sim 60$   & $A\sim 80$  \\
\smallskip
              & $s=1.6$, 1/1.6  & $A\sim 9$    & $A\sim 16$   & $A\sim 20$  & $A\sim 745$  & --          \\

main          & $s=1.2$, 1/1.2  & $A\sim 6$    & $A\sim 11$   & $A\sim 20$  & $A\sim 55$   & $A\sim 100$  \\
sequence      & $s=1.4$, 1/1.4  & $A\sim 8$    & $A\sim 20$   & $A\sim 30$  & $A\sim 100$  & $A\sim 150$  \\
              & $s=1.6$, 1/1.6  & $A\sim 11$   & $A\sim 30$   & $A\sim 40$  & $A\sim 150$  & $A\sim 200$  \\
\enddata 
\tablecomments{ 
Critical magnifications at which transition from the regime of perturbations 
induced by planetary caustics into the one of perturbations induced by 
central caustics occur. We note that the critical magnifications are 
$\lesssim 20$ in many cases.
}
\end{deluxetable*}

\begin{figure}[bht]
\epsscale{1.15}
\plotone{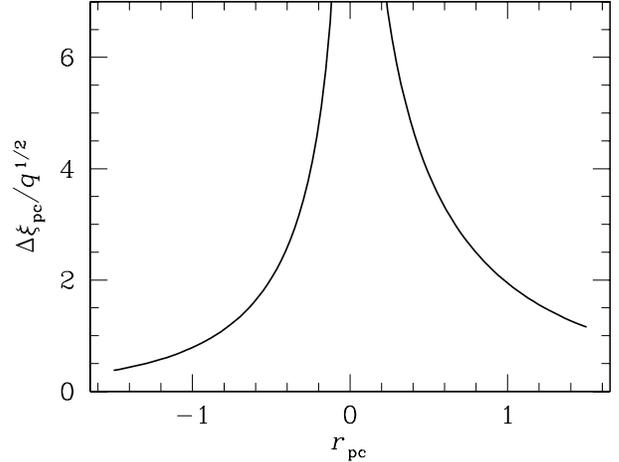}
\caption{\label{fig:four}
Variation of the size of the planetary caustic as a function of its location.  
The value $r_{\rm pc}$ represents the separation between the center of 
the planetary caustic and the primary lens star.  The sign of $r_{\rm pc}$ 
is positive when the caustic is on the planet side and vice versa.  We note 
that the caustic size at around $r_{\rm pc}$ is not presented because the 
analytic expression in eq.~(\ref{eq1}) is not valid in this region.  In 
addition, there is no distinction between the planetary and central caustics 
in this region.
}\end{figure}

From the variation of the probability, we find the following tendencies.  
First, we find that the probability increases with the increase of the 
lensing magnification.  This is consistent with the result of K.\ Horne 
(private communication).  This tendency is due to three factors.  First, 
the size of the planetary caustic increases as it is located closer to 
the primary star.  This can be seen in Figure~\ref{fig:four}, where we 
present the relation between the location of the planetary caustic and 
its size, which is obtained by using equations (\ref{eq2}) and (\ref{eq3}).  
Then, higher chance of planetary perturbation is expected when the source 
is located closer to the primary during which the lensing magnification 
is high.  Second, perturbation regions of the same size cover a larger 
range of angle as the planetary caustic moves closer to the lens.  This 
also contributes to the higher probability.  Third, the photometric 
precision improves with the increasing brightness of the source star due 
to lensing magnification.  As the photometric precision improves, it is 
easier to detect small deviations induced by planets.  The same reason 
can explain the considerable size of the perturbation region induced by 
central caustics.  Perturbations induced by the central caustics occur 
at high magnifications during which the photometric precision is high.  
As a result, despite much smaller size of the central caustic than that 
of the planetary caustic, the central perturbation region is considerable 
and can even be comparable to the perturbation region induced by the 
planetary caustic.  This can be seen in the detectability maps presented 
in Figure~\ref{fig:one}.

However, the probability does not continue to increase with the increase
of the magnification.  Instead, the probability drops off rapidly beyond a 
certain magnification.  This critical value corresponds to the magnification 
at which finite-source effect begins to wash out the planetary signal.
In Table~\ref{table:one}, we present the cases where finite source effect 
limits planet detections.  As a result, detections of planets with low 
mass ratios would be difficult for events involved with giant source stars 
with magnifications $A\gtrsim 70$.  We note that the finite source effect 
also limits the maximum magnifications of events and thus the curves in 
Figure~\ref{fig:three} discontinue at a certain value.

Second, as the magnification increases, the probability of detecting 
planetary signal increases with two dramatically different rates of 
$dP/d \log A$.  We find that this abrupt change of $dP/d \log A$ occurs 
due to the transition from the regime of perturbations induced by planetary 
caustics into the one of perturbations induced by central caustics.  The 
perturbation region induced by the central caustic forms around the primary 
lens and thus the probability becomes very high once the source star is in 
the central perturbation regime.  The boost of the increase rate occurs at 
different magnifications depending on the planetary parameters and the 
types of involved source stars.  The critical magnification becomes lower 
as the mass ratio of the planet increases and the separation of the planet 
approaches the Einstein ring radius.  In Table~\ref{table:two}, we present 
these critical magnifications.  An important finding to be noted is that 
the critical magnification occurs at moderate magnifications of $\lesssim 
20$ for a significant fraction of events caused by planetary systems with 
planets located in the lensing zone.  This implies that probability of 
detecting planetary signal can be high even for events with moderate 
magnifications.

Third, the probability is higher for events involved with brighter source 
stars.  This is because of the improved photometric precision with the 
increase of the source brightness.  The difference in the probability
depending on the source type is especially important at low magnifications.  
For example, the probabilities at a magnification of $A=5$ for events 
caused by a common planetary system with $q=10^{-3}$ and $s=1.2$ but 
associated with different source stars of giant, clump giant, and 
main-sequence are $P\sim 20\%$, 10\%, and 1\%, respectively.  In the 
absence of high magnification events, therefore, the second prime candidate 
event for follow-up observation is the one involved with brightest source 
star.  As the magnification further increases and once the source star 
enters the central perturbation region, the difference becomes less 
important.

\section{Summary and Conclusion}

For the purpose of providing useful criteria in the selection of target 
events preferable for planetary lensing follow-up observations, we 
investigated the variation of the probability of detecting planetary 
lensing signals depending on the observables of the lensing magnification 
and source brightness.   From this investigation, we found consistent 
result from previous studies that the probability increases with the 
increase of the lensing magnification due to the improvement of the 
photometric precision combined with the expansion of the perturbation 
region.  The increase rate of the probability is boosted at a certain 
magnification at which perturbation caused by the central caustic begins 
to occur.  We found that this boost occurs at moderate magnifications of 
$A\lesssim 20$ for a significant fraction of events caused by planetary 
systems with planets located in the lensing zone, implying that 
probabilities can be high even for events with moderate magnifications.  
The probability increases with the increase of the source star brightness.
We found that the probability of events associated with source stars 
brighter than clump giants is not negligible even at magnifications as 
low as $A\sim 5$.  In the absence of rare prime target of very 
high-magnification events ($A\gtrsim 100$), we, therefore, recommend 
to observe events with brightest source stars and highest magnifications 
among the alerted events.  Due to the increase of the source size with 
the increase of the brightness, however, the probability rapidly drops 
off beyond a certain magnification.  As a result, detections of planets 
with low mass ratios ($q\lesssim 10^{-4}$) would be difficult for events 
involved with giant source stars with magnifications $A\gtrsim 70$.

\acknowledgments 

This work was supported by the Astrophysical Research Center for the
Structure and Evolution of the Cosmos (ARCSEC) of Korea Science and
Engineering Foundation (KOSEF) through Science Research Program (SRC)
program.

\end{document}